\documentclass{emulateapj}
\usepackage{apjfonts}
\lefthead{HAN} 
\righthead{CHANNELS OF PLANETARY LENSING}

\begin{document}
\title{Expansion of the Planet Detection Channels in Next-Generation 
Microlensing Surveys}

\author{Cheongho Han}
\affil{Program of Brain Korea 21, Department of Physics, Institute 
for Basic Science Research, Chungbuk National University, Chongju 
361-763, Korea;\\
cheongho@astroph.chungbuk.ac.kr}

% ==================================================================

%\submitted{Submitted to The Astrophysical Journal}

\begin{abstract}
We classify various types of planetary lensing signals and the 
channels of detecting them.  We estimate the relative frequencies 
of planet detections through the individual channels with special 
emphasis on the new channels to be additionally provided by future 
lensing experiments that will survey wide fields continuously at 
high cadence by using very large-format imaging cameras.  From this 
investigation, we find that the fraction of wide-separation planets 
that would be discovered through the new channels of detecting 
planetary signals as independent and repeating events would be 
substantial.  We estimate that the fraction of planets detectable 
through the new channels would comprise $\sim 15$ -- 30\% of all 
planets depending on the models of the planetary separation 
distribution and mass ratios of planets.  Considering that a 
significant fraction of planets might exist in the form of free-floating 
planets, the frequency of planets to be detected through the new channel 
would be even higher.  With the expansion of the channels of detecting 
planet, future lensing surveys will greatly expand the range of planets 
to be probed.
\end{abstract}


\section{Introduction}

Microlensing is one of the important techniques that can detect and 
characterize extrasolar planets.  The technique is important especially 
in the detections of low-mass planets and it is possible to detect 
Earth-mass planets from ground-based observations.  The capability of 
the microlensing technique has been demonstrated by the recent detections 
of four planets \citep{bond04, udalski05, beaulieu06, gould06}.  Among 
them, one (OGLE-2005-BLG-390Lb) is the lowest mass planet ever detected
among those orbiting normal stars.

Planetary lensing signal lasts a short period of time; several days 
for a Jupiter-mass planet and several hours for an Earth-mass planet.  
To achieve the observational frequency required for the detection of 
the short-lived planetary signal, current microlensing planet search 
experiments are being operated in an observational setup, where survey 
observations (e.g., OGLE: \citet{udalski03}, MOA: \citet{bond02a}) issue 
alerts of ongoing events and subsequent follow-up observations (e.g., 
PLANET: \citet{albrow01}, MicroFUN: \citet{dong06}) intensively monitor 
the alerted events.  Under this strategy, however, only planetary signals 
occurring during the lensing magnification of the source star can be 
effectively monitored.  These signals are produced by planets having 
projected separations from the star similar to the Einstein radius of 
the primary star.  As a result, only planets located in a narrow region 
of separation from the host star can be effectively detected under the 
current planetary lensing searches.

The range of planets detectable with the microlensing technique will 
be expanded with next-generation lensing experiments that will 
survey wide fields continuously at high cadence by using very 
large-format imaging cameras.  Several such surveys in space and on 
ground are being seriously considered.  {\it Microlensing Planet 
Finder} ({\it MPF}), that succeeds the original concept of {\it 
Galactic Exoplanet Survey Telescope} ({\it GEST}) \citep{bennett02}, 
is a space mission proposed to NASA's Discovery Program with the main 
goal of searching for a large sample of extrasolar planets by using 
gravitation lensing technique \citep{bennett04}.  The `Earth-Hunter' 
project is a ground-based microlensing survey that plans to achieve 
$\sim 10$ minute sampling by using a distributed network of multiple 
wide-field ($\sim 2^\circ\times2^\circ$) telescopes (A.\ Gould, private 
communication). Recently, the MOA group begin a strategy observing a 
fraction of their fields very frequently by using a recently 
upgraded 1.8 m telescope with a $2.2\ {\rm deg}^2$ field of view.
These surveys dispense with the alert/follow-up mode of searching for 
planets and instead simultaneously obtain densely and continuously 
sampled light curves of all microlensing events in the field-of-view.  
With these surveys, the efficiency of planet detection will greatly 
improve thanks to the enhanced monitoring frequency and continuous 
sampling.  In addition, the future lensing experiments will be able 
to open new channels of planet detections because source stars are 
monitored regardless of their magnifications.  In this paper, we 
classify various channels of detecting planetary lensing signals 
and estimate the relative frequencies of planet detections through the 
individual channels with special emphasis on the new channels that 
will be provided by future lensing surveys.

The format of the paper is as follows.  In \S\ 2, we describe basics 
of planetary microlensing.  In \S\ 3, we present various types of 
planetary lensing perturbations and classify the channels of detecting 
them.  In \S\ 4, we estimate the relative frequencies of detecting 
planets through the individual channels.   We discuss the importance
of the new planet detection channels to be provided by future lensing 
surveys.  We summarize the results and conclude in \S\ 5.

% ==========================================================================

\section{Basics of Planetary Microlensing}

Due to the existence of a planetary companion, description of the  
planetary lensing behavior requires formalism of binary lensing 
\citep{witt90, witt95}.  Because of the small mass ratio of the 
planet, the light curve of a planetary lensing event is well described 
by that of a single lens of the primary star for most of the event 
duration.  However, a short-duration perturbation can occur when the 
source star passes the region around the caustics. The caustics are 
important features of binary lensing and they represent the set of 
source positions at which the magnification of a point source becomes 
infinite.  The caustics of binary lensing form a single or multiple 
sets of closed curves where each of which is composed of concave 
curves (fold caustics) that meet at points (cusps).  For a planetary 
case, there exist two sets of disconnected caustics: `central' and 
`planetary' caustics.

The single central caustic is located close to the host star.  It 
has a wedge shape with four cusps, where two are located on the 
star-planet axis and the other two are located off the axis.  The 
size of the central caustic as measured by the separation between 
the on-axis cusps is represented by \citep{chung06}
\begin{equation}
\Delta\xi_{\rm cc} \sim {4q\over (s-s^{-1})^2},
\label{eq1}
\end{equation}
where $q$ is the planet/star mass ratio and $s$ is the star-planet 
separation normalized by the Einstein radius of the planetary lens 
system, $\theta_{\rm E}$.  Then the caustic size
%, which is directly 
%proportional to the cross-section of the planetary perturbation region, 
becomes maximum when $s\sim 1.0$.  In the limiting case of a very 
wide-separation planet ($s\gg 1.0$) and a close-in planet ($s\ll 1.0$), 
the caustic size decreases, respectively, as
\begin{equation}
\Delta\xi_{\rm cc} \propto
\cases{
s^{-2}  & for $s \gg 1.0$, \cr
s^2     & for $s \ll 1.0$.\cr }
\label{eq2}
\end{equation}
For a given mass ratio, a pair of central caustics with separations 
$s$ and $s^{-1}$ are identical to the first order of the approximation
where the planet-induced anomalies is treated as a perturbation 
\citep{dominik99, griest98,an05}.  The central caustic is always 
smaller than the planetary caustic.  Since the central caustic is 
located close to the primary lens, the perturbation induced by the 
central caustic always occurs near the peak of high-magnification 
events.

The planetary caustic is located away from the host star.  The center 
of the planetary caustic is located on the star-planet axis and the 
position vector to the center of the planetary caustic from the primary 
lens position is related to the lens-source separation vector, ${\bf s}$, 
by 
\begin{equation}
{\bf r}_{\rm pc}={\bf s}\left(1-{1 \over s^2}\right).
\label{eq3}
\end{equation}
Then, the planetary caustic is located on the planet side, i.e.\ 
${\rm sign} ({\bf r}_{\rm pc})= {\rm sign}({\bf s})$, when $s>1.0$, and 
on the opposite side, i.e.\ ${\rm sign} ({\bf r}_{\rm pc})=-{\rm sign}
({\bf s})$, when $s<1.0$.  When $s>1.0$, there exists a single planetary 
caustic and it has a diamond shape with four cusps.  When $s<1$, there 
are two caustics and each has a triangular shape with three cusps.  
The size of the planetary caustic is related to the planet parameters by
\begin{equation}
\Delta\xi_{\rm pc} \propto 
\cases{
q^{1/2}\left[ s (s^2-1)^{1/2} \right]^{-1}             & for $s > 1$,\cr
q^{1/2}(\kappa_0-\kappa_0^{-1}+\kappa_0s^{-2})\cos\theta_0  & for $s < 1$,\cr
}
\label{eq4}
\end{equation}
where $\kappa(\theta)=\left\{[\cos 2\theta\pm (s^4-\sin^2 2\theta)^{1/2}]
/(s^2-s^{-2})\right\}^{1/2}$, $\theta_0 = [\pi \pm \sin^{-1}(3^{1/2}s^2/2)]
/2$, and $\kappa_0=\kappa(\theta_0)$ \citep{han06}.  We note that the 
dependence of the planetary caustic size on the mass ratio is $\Delta 
\xi_{\rm pc}\propto q^{1/2}$, while the dependence of the central caustic 
size is $\Delta \xi_{\rm cc}\propto q$.  Therefore, the decay rate of 
the planetary caustic with the decrease of the planet mass is slower 
than that of the central caustic.  The planetary caustic is located 
within the Einstein ring of the primary star when the planet is located 
in the range of separation from the star of $0.6 \lesssim s \lesssim 1.6$.  
The size of the caustic is maximized when the planet is located in this 
range, and thus this range is called as the `lensing zone' \citep{gould92, 
griest98}.  In the limiting cases of planetary separation, the size of 
the planetary caustic deceases similar to the central caustic, i.e.
\begin{equation}
\Delta\xi_{\rm pc} \propto
\cases{
s^{-2}  & for $s \gg 1.0$, \cr
s^2     & for $s \ll 1.0$.\cr }
\label{eq5}
\end{equation}
In the limiting case of $s\ll 1.0$, the two lens components work as if 
they are a single lens.  In the limiting case of $s\gg 1.0$, on the other 
hand, the star and planet work as if they are two independent lenses.

% ==========================================================================

% Figure 1 
\begin{figure}[t]
\epsscale{1.2}
\plotone{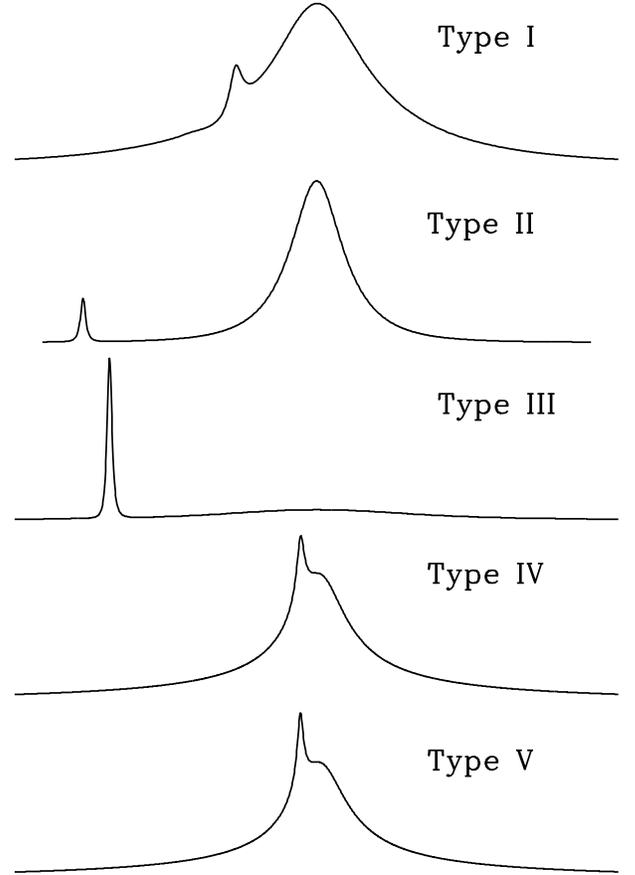}
\caption{\label{fig:one}
Schematic representation of the planetary signals with various types.
For the details of the classification scheme, see \S\ 3.
}\end{figure}

\section{Classification of Planet Detection Channels} 

Planetary lensing signal takes various forms depending on the 
characteristics of the planetary system, especially on the star-planet 
separation, and the source trajectory with respect to the positions 
of the star and planet.  In this section, we classify various types 
of planetary lensing signals and the channels of detecting them.

Type I perturbation shows up as a perturbation to the smooth light 
curve of the primary-induced lensing event (see the top panel in 
Figure~\ref{fig:one}).  This type of perturbation is produced by 
planets with projected star-planet separations similar to the Einstein 
radius of the primary star, i.e.\ $s\sim 1.0$.  As a result, the 
channel of detecting planets through the detection of type I perturbation 
is often referred as the `resonant channel'.  The resonant channel is 
the prime channel of detecting planets in current planetary lensing 
searches, which are based on survey/follow-up mode.  Depending on 
whether the perturbation is induced by the planetary or central 
caustic, the perturbation takes place on the side or near the peak 
of the light curve.

Type II perturbation is produced by a planet with a projected separation 
from the primary star substantially larger than the Einstein radius of 
the primary star ($s\gg 1.0$) and it occurs when the source trajectory 
passes both the effective magnification regions of the primary star 
and planet.  The planetary signal is the planet-induced lensing light 
curve itself that is well separated from the light curve of the primary 
(see the second panel in Figure~\ref{fig:one}).  Since two successive 
events are produced by the star and planet, respectively, this channel 
is often referred as the `repeating channel' \citep{distefano99}.  The 
planetary signal occurs long after or before the event induced by the 
primary star, and thus type II perturbation is difficult to be detected 
by the current planetary lensing searches monitoring only during the 
time of primary-induced lensing magnification.

Type III perturbation is produced also by a wide-separation planet,
but it occurs when the source trajectory passes only the effective 
magnification region of the planet.  Then, the planetary signal is the
independent lensing light curve produced by the planet itself (see the 
third panel in Figure~\ref{fig:one}).  We refer this channel of planet 
detection as the `independent channel'.  
%Another population of planets 
%that can be detected through the independent channel are free-floating 
%planets \citep{bennett02, han04}.  
Type III perturbation is also difficult 
to be detected by the current planetary lensing searches.

% Figure 2 
\begin{figure*}[ht]
\epsscale{0.80}
%\begin{figure}[t]
%\epsscale{1.25}
\plotone{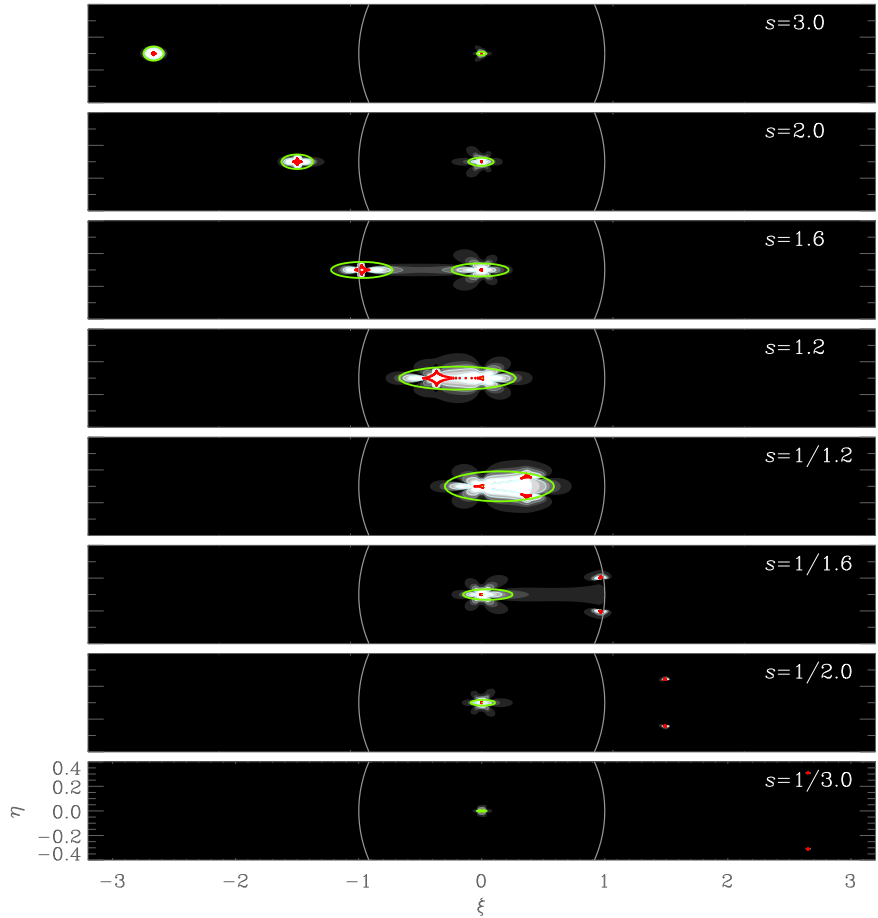}
\caption{\label{fig:two}
Maps of planetary signal detectability as a function of the source 
position for planetary systems with various projected star-planet 
separation.  The detectability is quantified as the ratio of the 
fractional deviation of the planetary lensing light curve from that 
of the single lensing event of the primary star to the photometric 
precision.  The maps are centered at the position of the primary lens 
star and the planet is located on the left.  The green (green curve) 
is drawn at the level of $D=3$.  The solid circle centered at the 
primary star in each map represents the Einstein ring.  The closed 
figures drawn by red curves are the caustics.  All lengths are in 
units of the Einstein radius of the planetary system and $\xi$ and 
$\eta$ represent the coordinates that are parallel and normal to the 
star-planet axis, respectively.  Each planetary perturbation region 
is approximated by an elliptical region and the green ellipse represents 
its boundary.  The planetary lens system has a common planet/star mass 
ratio of $q=3\times 10^{-3}$.  
}\end{figure*}
%}\end{figure}

Wide-separation planets can also be detected through another channel.
Type IV perturbation is produced by a wide-separation planet and it 
occurs when the source trajectory passes very close to the primary 
star.  The planetary signal is then a brief perturbation near the peak 
of the high-magnification lensing light curve produced by the primary 
star (see the fourth panel in Figure~\ref{fig:one}).  We refer this 
channel as the `wide central channel'.  High-magnification events 
are of highest-priority in the current microlensing follow-up 
observations because of their high sensitivity to planets
\citep{griest98, albrow00, bond02b, rattenbury02, abe04, yoo04, 
dong06} and thus type IV perturbation can be detected from current 
follow-up observations.

Type IV perturbation is produced by a planet with a star-planet 
separation substantially smaller than the Einstein radius of the 
primary star and it occurs when the source trajectory passes very 
close to the primary star.  The planetary signal is very similar to 
the type IV perturbation (see the fourth panel in Figure~\ref{fig:one}) 
due to the $s\leftrightarrow s^{-1}$ symmetry of the central caustic 
\citep{dominik99, an05}.  As a result, distinguishing the two types 
of perturbation is often difficult, especially for perturbations 
induced by low-mass planets with mass ratios $q\lesssim 10^{-4}$ 
\citep{chung06}.

% ==========================================================================

\section{Frequency of the Individual Channels}

In the previous section, we presented various types of planetary 
perturbations and the channels of detecting them.  Then, what would 
be the relative frequencies of detecting planets through the individual 
channels from future lensing surveys?  In this section, we answer to 
this question.

The rate of planet detection is proportional to the cross-section of 
the planetary perturbation region, $\sigma$.  We, therefore, estimate 
the relative frequencies by computing the cross-sections of the planetary 
perturbation regions of the corresponding types.  We proceed our 
computation according to the following procedure.
\begin{enumerate}
\item
First, we make maps of perturbation induced by planets with various 
separations and mass ratios.
\item
Second, we classify the types of perturbation based on the planetary 
separation and the location of the perturbation regions.  We then 
estimate the average cross-sections of the perturbation regions belonging 
to the individual categories based on the constructed perturbation maps.  
\item
Finally, we compute the relative frequencies of detecting planets through 
the individual channels by convolving the cross-sections with the model 
distributions of planetary separation.  
\end{enumerate}
The details of the individual processes are described in the following
subsections.

\subsection{Maps of Planetary Signal Detectability}

The map of planetary perturbation represents the region of planet-induced 
lensing perturbations as a function of source position.  The quantity 
that has been often used to represent the perturbation region is the 
`fractional deviation' of the planetary lensing light curve from that 
of the single lensing event of the primary star, i.e.,
\begin{equation}
\epsilon = {A-A_0 \over A_0},
\label{eq6}
\end{equation}
where $A$ and $A_0$ represent the lensing magnifications with and 
without the planet, respectively.  With this quantity, however, one
cannot consider the variation of the photometric precision depending 
on the lensing magnification and source brightness.  To consider this,
we construct the map of perturbation with the quantity 
defined as the ratio of the fractional deviation, $\epsilon$, to the 
photometric precision, $\sigma_\nu$, i.e, 
\begin{equation}
{\cal D}={\left\vert \epsilon\right\vert \over \sigma_\nu};\qquad
\sigma_\nu = { ( AF_{\nu,{\rm S}}+F_{\nu,{\rm B}})^{1/2} 
\over (A-1)F_{\nu,{\rm S}}},
\label{eq7}
\end{equation}
where $F_{\nu,{\rm S}}$ and $F_{\nu,{\rm B}}$ represent the photon 
counts from the source star and blended background stars, respectively.  
Under this definition of the planetary perturbation, ${\cal D}=1$ 
implies that the planetary signal is equivalent to the photometric 
precision.  Hereafter, we refer the quantity ${\cal D}$ as the 
`detectability'.

To construct the map of detectability, we choose a representative 
Galactic bulge event.  Following the result of simulations of Galactic 
bulge events, e.g.\ \citet{han03, han05}, we choose a representative 
event as the one produced by a lens with the primary lens mass of 
$m=0.3\ M_\odot$ and the distances to the lens and source of $D_{\rm L}=6$ 
kpc and $D_{\rm S}=8$ kpc, respectively.  For the observational condition, 
we take the space-based lensing survey by using the {\it MPF} mission as 
a reference experiment.  The prime target source stars to be monitored 
by the {\it MPF} survey are main-sequence stars and thus we choose a 
main-sequence source star with an $I$-band absolute magnitude of $M_I=4.8$, 
which corresponds to a K0 star.  With the assumed amount of extinction 
toward the Galactic bulge field of $A_I=1.0$, this correspond to the 
apparent magnitude of $I=20.3$.  Following the specification of the 
{\it MPF} mission, we assume that the photon acquisition rate is 13 
photons per second for an $I=22$ star and photometry is done on each 
combined image with an exposure time $t_{\rm exp}=10$ minutes.  We 
assume that blending is not important due to high resolution from 
space-based observation.  Finite size of the source star might affect 
the planet detectability \citep{bennett96}.  However, the populations 
of planets of our interest are the ones to be detected through the new 
channels (i.e., repeating and independent channels) and most of them 
would be giant planets because of their larger cross-sections.
Considering that the angular size of the source star is a few percent 
of the angular Einstein radius of a giant planet, finite size of the 
source star has little effect on the planet detectability.  We therefore 
do not consider finite-source effect in our analysis.

Figure~\ref{fig:two} shows constructed maps of detectability for some 
example planetary systems.  The maps are centered at the position of 
the primary lens star and the planet is located on the left.  The 
contour (green curve) is drawn at the level of $D=3$, within which the 
planetary signal is detected with a $3\sigma$ confidence level.  The 
solid circle centered at the primary star in each map represents the 
Einstein ring and the close figure drawn by red curves represent the 
caustics.  All lengths are normalized in units of the Einstein radius 
corresponding to the mass of the primary star, $\theta_{\rm E}$.

% Figure 3 
\begin{figure}[t]
\epsscale{1.20}
\plotone{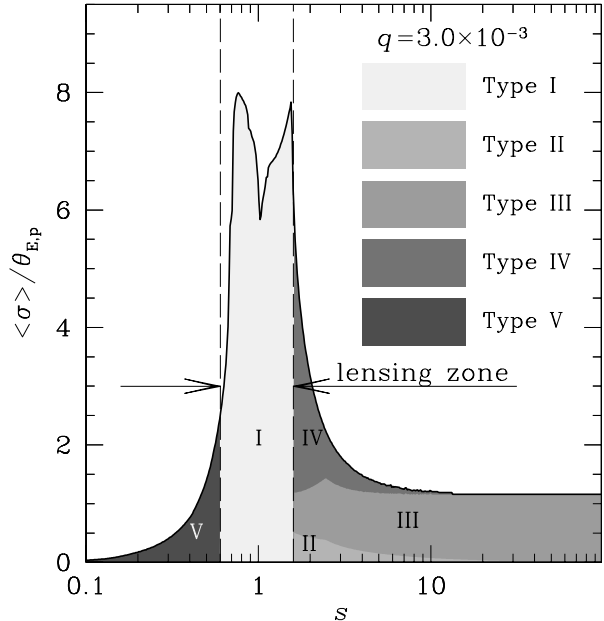}
\caption{\label{fig:three}
Cross-section of planetary perturbation region as a function of the
normalized star-planet separation.  The segments marked by different 
tones of shade under the curve represent the types of the related
planetary perturbations.  We note that the cross-section is normalized 
by the angular Einstein radius of the planet.  The assumed planet/star 
mass ratio is $q=3\times 10^{-3}$.
}\end{figure}

\subsection{Cross-Section of Perturbation}

With the constructed maps of perturbation, we then estimate the 
average cross-sections of the perturbation regions of the individual 
types.  For this, we classify the types of perturbation based on the 
planetary separation and the location of the perturbation region.

The followings are the criteria for the classification.  First, we 
classify all perturbations induced by planets located in the lensing 
zone ($0.6 \leq s \le 1.6$) into type I.  If a perturbation is induced 
by a wide-separation ($s>1.6$) planet, the perturbation region is 
divided into two parts; one around the planetary caustic and the other 
around the central caustic.  If the perturbation is caused by the 
central caustic, it is classified into type IV perturbation.  Among 
the perturbations caused by the planetary caustics of wide-separation 
planets, the fraction of the type II perturbation is geometrically 
estimated as $\sin(2/\pi) \sin^{-1}( u_{\rm th}/s)$, where $u_{\rm th}$ 
is the radius of the effective lensing magnification region of the 
primary star.  We adopt $u_{\rm th}=1.5$.  Then the rest of perturbations 
induced by the planetary caustics of wide-separation planets are 
classified into type III.  Finally, perturbations induced by planets 
with separations $s<0.6$ are classified into type V.

Once the types of the individual perturbation regions are determined, 
we then estimate the average cross-sections of the perturbation regions.  
A straightforward approach to estimating the cross-section would be 
first drawing many light curves resulting from source trajectories with 
various combinations of the distance to the trajectory from the center 
of the perturbation region and orientation angles, then checking the 
detectability of the planet-induced perturbations for the individual 
light curves, and finally estimating the cross-section as an angle-averaged 
value.  However, this requires a large amount of computation time.  
Fortunately, the perturbation region is confined around caustics and 
its boundary is approximated by an ellipse.  We therefore estimate the 
cross-section by approximating the perturbation region as an elliptical 
region.  With this approximation, the cross-section of the perturbation 
region is estimated as the angle-averaged cross-section of the ellipse, 
i.e.\ 
\begin{equation}
\langle\sigma\rangle = {1\over \pi}\int_0^\pi 
\left( a_{\rm p}^2\sin^2 \theta + b_{\rm p}^2\cos^2\theta \right)^{1/2} 
d\theta = {2\over \pi} a_{\rm p} E(e),
\label{eq8}
\end{equation}
where $a_{\rm p}$ and $b_{\rm p}$ are the semimajor and semiminor axes 
of the elliptical boundary of the perturbation region, $e=(1-b_{\rm p}^2
/a_{\rm p}^2)^{1/2}$ is the eccentricity of the ellipse, and $E$ 
represents the complete elliptical integral of the second kind.  We 
determine the semimajor and semiminor axes of the ellipse as the widths 
of the perturbation region enclosed by the detectability contour with a 
level of $D=3$ along and normal to the star-planet axis, respectively.  
If the perturbation region is composed of multiple segments, we approximate 
the individual segments with different ellipses.  In Figure 2, we present 
the elliptical boundaries of perturbation regions (green curves) on the 
top of the detectability map.

Figure~\ref{fig:three} shows the determined cross-section of the 
planetary perturbation region as a function of the normalized star-planet 
separation for a planetary lens with a mass ratio $q=3\times 10^{-3}$, 
which corresponds to a Jupiter-mass planet around a primary star with a
mass $m=0.3\ M_\odot$.  We note that the cross-section is normalized by 
the angular Einstein radius corresponding to the mass of the planet, 
$\theta_{\rm E,p}=q^{1/2} \theta_{\rm E}$.  The segments marked by 
different tones of shade under the curve represent the types of the 
related perturbations.  From the figure, one finds that the cross-section 
vanishes in the limiting case of $s\rightarrow 0$.  This is because 
the star and planet work as if they are a single lens in this limit.  
In the limiting case of the other end ($s\rightarrow \infty$), on the 
other hand, the planet acts as an independent lens and thus the 
cross-section converges into the value corresponding to the cross-section 
of the effective magnification region of the planet.

\begin{deluxetable*}{ccrrrrr}
\tablecaption{Relative Frequency\label{table:one}}
\tablewidth{0pt}
\tablehead{
\multicolumn{1}{c}{planetary separation} &
\multicolumn{1}{c}{planet/star} &
\multicolumn{5}{c}{types of planetary perturbation} \\
\multicolumn{1}{c}{distribution model} &
\multicolumn{1}{c}{mass ratio} &
\multicolumn{1}{c}{type I} &
\multicolumn{1}{c}{type II} &
\multicolumn{1}{c}{type III} &
\multicolumn{1}{c}{type IV} &
\multicolumn{1}{c}{type V} 
}
\startdata
$\alpha=1.0$ & $q=5.0\times 10^{-3}$ & 48.3\%  & 3.8\%  & 26.5\%  & 12.8\%  &  8.6\% \\
             & $q=1.0\times 10^{-3}$ & 56.8\%  & 4.1\%  & 25.9\%  &  6.9\%  &  6.2\% \\
             & $q=5.0\times 10^{-4}$ & 59.4\%  & 4.2\%  & 25.6\%  &  5.4\%  &  5.3\% \\
\smallskip
             & $q=1.0\times 10^{-4}$ & 64.3\%  & 4.0\%  & 24.3\%  &  3.7\%  &  3.7\% \\

$\alpha=1.5$ & $q=5.0\times 10^{-3}$ & 58.2\%  & 2.5\%  & 12.7\%  & 10.0\%  & 16.6\% \\
             & $q=1.0\times 10^{-3}$ & 67.2\%  & 2.9\%  & 12.7\%  &  5.3\%  & 11.9\% \\
             & $q=5.0\times 10^{-4}$ & 69.9\%  & 3.0\%  & 12.7\%  &  4.2\%  & 10.0\% \\
             & $q=1.0\times 10^{-4}$ & 75.0\%  & 2.8\%  & 12.1\%  &  2.9\%  &  7.2\% 

%$\alpha=2.0$ & $5.0\times 10^{-3}$ & 58.0\%  & 1.4\%  &  5.6\%  &  6.8\%  & 28.2\% \\
%             & $1.0\times 10^{-3}$ & 68.0\%  & 1.7\%  &  6.0\%  &  3.6\%  & 20.6\% \\
%             & $5.0\times 10^{-4}$ & 71.4\%  & 1.8\%  &  6.1\%  &  2.8\%  & 17.8\% \\
%             & $1.0\times 10^{-4}$ & 77.1\%  & 1.8\%  &  6.1\%  &  2.0\%  & 13.0\% 
\enddata
\tablecomments{
Relative frequencies of detecting planets through various channels.
The frequencies are estimated under the assumption that the star-planet 
separation follows a power-law distribution of $dN/da \propto a^{-\alpha}$,
where $\alpha$ represents the power of the distribution.
}
\end{deluxetable*}

\subsection{Relative Frequencies}

Once the average cross-sections of the individual types of perturbation
is computed, we then estimate the relative frequencies of detecting planets
through the individual channels of planet detections.  This is done 
by convolving the cross-section with model distributions of planetary 
separation.

We model the distribution of star-planet separation as a power-law
function of the form
\begin{equation}
{dN\over da} \propto a^{-\alpha},
\label{eq9}
\end{equation}
where $a$ is the semimajor axis of the planet orbit.  There is little 
consensus about the power of the distribution.  From the analysis of 
observed extrasolar planets detected by radial velocity surveys, 
\citet{tabachnik02} claimed $\alpha\sim 1$.  On the other hand, 
\citet{hayashi95} claimed that the surface density distribution of the 
minimum mass solar nebula is well described with $\alpha\sim 1.5$.  
%\citet{kuchner04} argued that multi-planet extrasolar planetary 
%systems indicate $\alpha\sim 2.0$.  
We, therefore, test two different powers of $\alpha=1.0$ and 1.5.
We note that larger absolute value of the power implies that planets 
are populated in the inner region.  Then the fraction of planetary 
events detectable through the type I perturbation increases while the 
fractions through the type II, III, and IV decrease.  We assume that 
planets are distributed up to a distance of 100 AU.  
%We note that the 
%fraction of planets detectable through the independent channel depends 
%critically on the choice of this limit.  
Once the semimajor axis of 
the planet orbit is determined, the projected star-planet separation 
is determined under the assumption of a circular orbit and random 
orientation of the orbital plane.  The projected separation is related 
to the intrinsic separation by
\begin{equation}
\tilde{a} = a(\sin^2 i \cos^2 \varphi + \cos^2 i)^{1/2},
\label{eq10}
\end{equation}
where $i$ is the inclination angle of the orbital plane and $\varphi$ 
is the phase of the planet on the orbital plane.

In Table~\ref{table:one}, we present the relative frequencies of 
detecting planets through the individual channels.  From the table, we 
find that the frequency of detecting planets through the new channels 
to be provided by future lensing surveys would be substantial.  We 
estimate that the fraction of planets detectable through the independent 
and repeating channels would comprise $\sim 15$ -- 30\% of all planets 
depending on the models of the planetary separation distribution and 
mass ratios of planets.  Considering that the total number of planets 
expected to be detected from five-year lensing surveys in space would 
be several thousands \citep{bennett04}, the number of planets detectable 
through the new channels would be of the order of hundred and can reach 
up to a thousand.  We note that the estimation in Table~\ref{table:one} 
is based only on planets bound to primary stars.

The new channels to be provided by future lensing surveys are important
for better understanding of planet formation and evolution processes.
Planets located $\gtrsim 5$ AU from host stars could not have been  
detected by any of the methods currently being used for planet 
searches.  Being able to detect planets in this range, therefore,
microlensing method would provide complete sample of planets.  Another 
population of planets that can be detected through the new channels 
are free-floating planets \citep{bennett02, han04}.  It is believed 
that a good fraction of planets have been ejected from their planetary 
systems during or after the epoch of planet formation \citep{zinnecker01}.
Another possible origin of these planets would be the accretion of gas 
similar to star formation process \citep{boss01}.  Since these planets 
were not included in our analysis, the relative frequency of planet 
detection through the new channels would be even larger if these planets 
are common.

\section{Conclusion}

We classified various types of planetary lensing signals and the 
channels of detecting them.  We estimated the relative frequencies 
of planet detections through the individual channels with special 
emphasis on the new channels that will be additionally provided by 
future lensing surveys.  From this investigation, we found that the 
fraction of wide-separation planets that would be discovered through 
the new channels of detecting planetary signals as independent and 
repeating events would be substantial.  We estimated that the fraction 
of planets detectable through the new channels would comprise $\sim 15$ 
-- 30\% of all planets depending on the models of the planetary separation 
distribution and mass ratios of planets.  Considering that a significant 
fraction of planets might exist in the form of free-floating planets, 
the frequency of planets to be detected through the new channel would 
be even higher.  We, therefore, demonstrate that future lensing surveys 
will greatly expand the range of planets to be probed.

\acknowledgments 
This work was supported by the grant (C00072) of the Korea Research
Foundation.

\end{document}